\documentclass[10pt,letterpaper]{article}
\usepackage[top=0.85in,left=1.7in,footskip=0.75in,marginparwidth=2in]{geometry}

\usepackage[utf8]{inputenc}

\usepackage{cite}
\usepackage[justify]{ragged2e}
\usepackage{nameref,hyperref}

\usepackage[right]{lineno}
\usepackage[T1]{fontenc}
\usepackage{float}
\usepackage{amssymb}
\usepackage{amsmath}
\usepackage{mathtools}
\usepackage[makeroom]{cancel}
\usepackage{xparse}
\usepackage{tikz}
\usetikzlibrary{matrix,backgrounds}
\pgfdeclarelayer{myback}
\pgfsetlayers{myback,background,main}
\usepackage[justify]{ragged2e}
\tikzset{mycolor/.style = {line width=1bp,color=#1}}%
\tikzset{myfillcolor/.style = {draw,fill=#1}}%

\NewDocumentCommand{\highlight}{O{blue!40} m m}{%
\draw[mycolor=#1] (#2.north west)rectangle (#3.south east);
}

\NewDocumentCommand{\fhighlight}{O{blue!40} m m}{%
\draw[myfillcolor=#1] (#2.north west)rectangle (#3.south east);
}

\usepackage{microtype}
\DisableLigatures[f]{encoding = *, family = * }

\usepackage{changepage}

\usepackage[aboveskip=1pt,labelfont=bf,labelsep=period,singlelinecheck=off]{caption}

\makeatletter
\renewcommand{\@biblabel}[1]{\quad#1.}
\makeatother

\usepackage{lastpage,fancyhdr,graphicx}
\usepackage{epstopdf}
\fancyheadoffset[L]{2.25in}
\fancyfootoffset[L]{2.25in}

\usepackage{color}

\definecolor{Gray}{gray}{.25}

\usepackage{graphicx}

\usepackage{sidecap}

\usepackage{wrapfig}
\usepackage[pscoord]{eso-pic}
\usepackage[fulladjust]{marginnote}
\reversemarginpar

\begin{document}
\vspace*{0.35in}

\begin{flushleft}
{\Large
\textbf\newline{The Impact of Time Delay on Mutant Fixation in Evolutionary Games}
}
\newline
\\
Javad Mohamadichamgavi\textsuperscript{1,*},
Mark Broom\textsuperscript{2}
\\
\bigskip
\bf{1} Institute of Applied Mathematics and Mechanics, University of Warsaw, ul. Banacha 2, 02-097 Warsaw, Poland
\\
\bf{2} City, University of London, Northampton Square, London, EC1V 0HB, UK
\\
\bigskip
* Jmohamadi@mimuw.edu.pl

\end{flushleft}

\section*{Abstract}
Evolutionary game theory examines how strategies spread and persist in populations through reproduction and imitation based on their fitness. Traditionally, models assume instantaneous dynamics where fitness depends on the current population state. However, some real-world processes unfold over time, with outcomes emerging from history. This motivates incorporating time delays into evolutionary game models, where fitness relies on the past. We study the impact of time delays on mutant fixation in a Moran Birth-death process with two strategies in a well-mixed population. At each time step of the process, an individual reproduces proportionally to fitness coming from the past. We model this as an absorbing Markov chain, allowing computational calculation of the fixation probability and time. We focus on three important games: the Stag-Hunt, Snowdrift, and Prisoner's Dilemma. We will show time delays reduce the fixation probability in the Stag-Hunt and the Prisoner's Dilemma but increase it in the Snowdrift. For the Stag-Hunt and the Prisoner's Dilemma, time delays lengthen the fixation time until a critical point, then reduce it. The Snowdrift exhibits the opposite trend.

\section{Introduction}
Evolutionary Game Theory (EGT) provides a framework for understanding how effective strategies spread through imitation or reproduction and persist in systems ranging from biology to economics\cite{smith1982evolution,hofbauer1998evolutionary,szabo2007evolutionary,nowak2006evolutionary,gintis2000game,broom2022game}. The theory gives insights into phenomena like cooperation, competition, and diversity in populations\cite{axelrod1981evolution,poundstone1993prisoner,skyrms2004stag,moyano2009evolving,hardin1968tragedy,hummert2014evolutionary,sachs2006cooperation}. It examines how strategies or types of individuals change in frequency in a population based on their fitness from interactions. Strategies that produce higher fitnesses will tend to spread and become more common as individuals reproduce or copy them. Traditional EGT models have primarily focused on well-mixed, infinitely large populations, where individuals interact at random and strategy frequencies evolve deterministically according to replicator dynamics or other differential equation models \cite{hofbauer1998evolutionary,taylor1978evolutionary,nowak2006evolutionary,szabo2007evolutionary}. However, real-world populations are typically finite and structured. To address these limitations, recent work has incorporated stochastic processes, such as the Moran process, into EGT models to study evolutionary dynamics in finite populations. \cite{nowak2004emergence,taylor2004evolutionary,imhof2005evolutionary,imhof2006evolutionary,nowak2004evolutionary,schaffer1988evolutionarily,komarova2003language,wild2004fitness,traulsen2006evolution,claussen2008cyclic,ohtsuki2006evolutionary,nowak2006evolutionary,nowak2004emergence}.

The classical concept of an Evolutionarily Stable Strategy (ESS) does not fully capture evolutionary dynamics in finite populations. A true ESS must now satisfy two criteria, it must resist invasion by rare mutant strategies, and its probability of replacement by a rare mutant must be lower than under neutral drift\cite{nowak2004evolutionary,taylor2004evolutionary,nowak2004emergence,schaffer1988evolutionarily}. The first requirement provides an equilibrium condition, while the second provides a stability condition. Selection opposes invasion when mutants cannot attain higher fitness than resident individuals in a pure population. It also opposes fixation when mutants have no better chance of reaching fixation than neutral mutations. Satisfying both criteria helps ensure a strategy is robust and maintains equilibrium\cite{broom2022game}. Therefore the study of fixation probabilities became a crucial component of comprehending EGT.

The Moran process is a classic model for studying evolution in finite populations\cite{moran1958random}. In this stochastic process, a population of $N$ individuals with different types undergoes birth and death events over discrete time steps. At each step, one individual is selected to reproduce and its offspring replaces another chosen individual\cite{allen2010introduction}. The individual's fitness can influence birth and death events in this process. In general, individuals reproduce with a probability proportional to their fitness and replace another individual at random (Bd). We note that this process can also happen in other ways, and there are a number of different dynamics used. Fitness can be constant or frequency-dependent based on game interactions between individuals\cite{nowak2004emergence,taylor2004evolutionary}. Typically the population is assumed well-mixed so that all individuals have the chance to be replaced by others. In this process, we track the fate of any emergent mutant, assessing the probability and the time that this single novel mutant ultimately fixes in the population (fixation probability and time). Also monitoring the trajectory of the mutant through the population provides insight into the dynamics of how novel mutations establish and propagate themselves in populations. 

The fixation probability and time have been extensively studied across various models\cite{de2019fixation,traulsen2006stochastic,traulsen2007pairwise,broom2010evolutionary,allen2021fixation,fudenberg2006evolutionary,antal2006fixation,sample2017limits,pires2022more,mobilia2010fixation}. Previous work has focused on how factors like population size\cite{de2019fixation,antal2006fixation,sample2017limits,pires2022more}, environmental fluctuations\cite{assaf2010large,czuppon2018fixation,whitlock2003fixation,ashcroft2014fixation}, and population structure \cite{ohtsuki2006simple,broom2008analysis,allen2014games,lieberman2005evolutionary,shakarian2012review,hindersin2016exact,shakarian2013novel,hindersin2014counterintuitive,hajihashemi2019fixation,mohamadichamgavi2023effect} influence mutant fixation. However, most of these studies assumed that fitness is determined instantaneously based on the current population state and payoffs. This assumption may not always be realistic, as biological and social processes often involve temporal delays.\cite{wu2016stochastic,milton2015time}. It is therefore natural for EGT models to incorporate time delays, where fitness depends on history.
The effects of time delays have been explored in the context of replicator dynamics for games with an interior stable equilibrium (ESS)\cite{yi1997effect,alboszta2004stability,mikekisz2021evolution,iijima2012delayed,ben2018discrete,wang2023evolutionary,miekisz2011stochasticity,wesson2016hopf,hu2019stability,wettergren2023replicator,bodnar2020three}. In \cite{yi1997effect}, the authors studied a model where individuals at time t imitate strategies that had higher average payoffs at time $t-\tau$ for some delay $\tau$. They showed the interior fixed point is locally asymptotically stable for small $\tau$, but becomes unstable for large $\tau$, leading to oscillations. In \cite{alboszta2004stability}, individuals are born after their parents play and receive payoffs. Two coupled equations governed strategy frequency and population size. Analysis showed oscillations are not possible, with the original fixed point globally asymptotically stable regardless of the time delay. In \cite{mikekisz2021evolution}, the authors modified the above model by allowing time delays to vary based on the strategies used by individuals. Under this framework, they observe novel behaviour where the equilibrium states become dependent on the delays. Most previous work only considered replicator dynamics and the impact of delays on equilibrium stability. Less attention has been given to finite populations and the effect of delays on the fixation process. 

Our study looks into the influence of time delays on the fixation process in evolutionary games. We explore the impact of time delays on the fixation probability and fixation time in a well-mixed population comprising two strategies, examining a generalized Moran Birth-death process, where the fitness of both strategies at time step $t$ hinges on the fitness values from $\tau$ steps previously. By constructing an absorbing Markov chain, whose states are defined by $\tau+1$ indices, $\tau$ indices representing the history, and one the current mutant abundance, we determine the transition matrix, enabling us to analyze the fixation probability and fixation time for various game types.

The paper is structured as follows. Section \ref{moran} reviews the Moran process definition and basic properties for two-strategy games. Section \ref{farmwork} introduces a general framework to calculate absorption probabilities and times in Markov chains related to the process. Section \ref{model} presents our model incorporating time delays into the Bd Moran process and defines a new associated Markov chain. Section \ref{results} details the results of our model and the effects of time delays on the fixation probability and time across different games. The paper finishes with a discussion of the key findings and a summary of the main results. 
\section{Moran Process}\label{moran}
We consider the evolutionary dynamics of individuals adopting two distinct strategies, A and B within a well-mixed population, the game being played characterized by the payoff matrix \cite{taylor2004evolutionary}:
\begin{center}
\begin{tabular}{l|ll}
  & A & B \\ \hline
A & a & b \\
B & c & d
\end{tabular}.
\end{center}
When a player with strategy A interacts with an A player (a B player), they receive a payoff of a (b). On the other hand, a B player receives a payoff of c (d) from an A (B) player. Consequently, the average payoffs when there are $i$ individuals with strategy A reads as follows:
\begin{equation}
\begin{split}
\pi_A(i)=\dfrac{(i-1)a+(N-i)b}{(N-1)}, \\
\pi_B(i)=\dfrac{ic+(N-i-1)d}{(N-1)}, \hspace{8 pt} 
\end{split}
\label{1}
\end{equation}
where self-interactions are excluded. The corresponding average fitnesses are proportional to:
\begin{equation}
\begin{split}
f_i = 1 - w + w \pi_A(i) , \\
g_i = 1 - w + w \pi_B(i) .
\end{split}
\label{efitnessfunction}
\end{equation}

The variable $w$ represents the intensity of selection, indicating how much the payoff of individuals affects fitness\cite{nowak2004emergence}. When $w$ approaches zero, each individual's payoff makes a minimal contribution to overall fitness, leading to what is known as weak selection. When $w=0$, all individuals possess the same level of the fitness, resulting in neutral drift. Conversely, as $w$ increases, the influence of payoffs on fitness grows and we have strong selection.

There are various update rules when individuals update their strategies during each time step of a process\cite{nowak2004emergence,ohtsuki2006simple,taylor2004evolutionary,pattni2017evolutionary,traulsen2006stochastic,hathcock2019fitness}. We use the Birth-death (Bd) process, where the fitness of individuals is considered in the Birth event which is a type of frequency-dependent Moran process\cite{nowak2004emergence,taylor2004evolutionary}. During each time step, an individual is chosen for reproduction based on their fitness. The newly born individual then replaces a randomly selected member of the population, ensuring that the total population size, $N$, remains constant. At each time step, $i$ can change by a maximum of one based on the following transition probabilities: 
\begin{equation}
\begin{split}
p_{i \longrightarrow  i+1}=\dfrac{if_i}{if_i+(N-i)g_i}\dfrac{N-i}{N-1}, \\
p_{i \longrightarrow  i-1}=\dfrac{(N-i)g_i}{if_i+(N-i)g_i}\dfrac{i}{N-1}.
\end{split}
\label{2}
\end{equation}

Here the objective is to examine the Bd process involving a population of residents with strategy B when introduced to a single mutant with strategy A.  This process ultimately results in one of two possible outcomes: the complete replacement of the population by the mutant, known as {\it{fixation}}, or the eradication of the mutant, referred to as {\it{extinction}}. Two quantities hold significant importance in this process, the fixation probability, which indicates the probability of the mutant becoming fixed in the population, and the fixation time, which represents the number of time steps required for the fixation\cite{antal2006fixation,hindersin2014counterintuitive,traulsen2009stochastic}. There are three main approaches used to analyze the fixation process: direct analytical solutions, numerical methods using transition matrices of Markov chains, and Monte Carlo simulations. Analytical solutions can be found for simple Moran processes without population structure or with regular structure, providing formulae for fixation times and probabilities. However, as models become more complex, researchers must study the dynamics using numerical techniques and Monte Carlo simulations.

\begin{figure}[!h]
    \centering
    \includegraphics[scale=0.25]{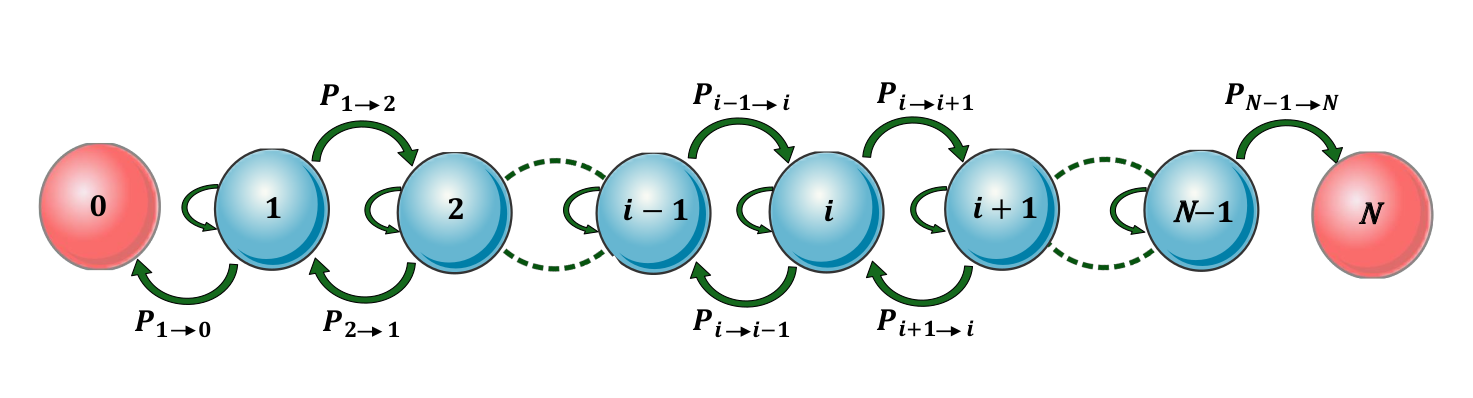}
    \caption{Absorbing Markov chain with $N+1$ states associated with the Bd Moran process on population with size $N$. Each state is determined by the number of mutants in the population $(i)$. States $0$ and $N$ are the absorbing states related to the extinction and fixation of mutants respectively.}
    \label{mark1}
\end{figure}

In general, we can consider an absorbing Markov chain associated with the process to investigate and compute the probability and time of fixation. Fig. \ref{mark1} provides a visual representation of the Markov chain. The state is determined by the number of mutants present in the population $(i)$. Essentially, the Markov chain represents a random walk on sites ranging from $0$ to $N$. Starting from state $i$, the walk eventually reaches either the absorbing states of $0$ (mutant extinction), or $N$ (mutant fixation). The transition matrix is defined as follows:

\begin{equation}
P_{i,j}=p_{i \longrightarrow  i+1} \delta_{i+1,j}+p_{i \longrightarrow  i-1} \delta_{i-1,j} + (1-p_{i \longrightarrow  i+1}-p_{i \longrightarrow  i-1}) \delta_{i,j}.
\label{3}
\end{equation}

The fixation probability $\Phi_{i,N}$ from state with $i$ initial mutants with strategy A under a Birth-death process can be calculated using the  recursive equations:

\begin{equation}
    \Phi_{i,N}=p_{i \longrightarrow  i-1}\Phi_{i-1,N}+(1-p_{i \longrightarrow  i-1}-p_{i \longrightarrow  i+1})\Phi_{i,N}+p_{i \longrightarrow  i+1}\Phi_{i+1,N}
    \label{efp1}
\end{equation}
Using this recursive relationship and the boundary conditions that $\Phi_{0,N}=0$ and $\Phi_{N,N}=1$, the fixation probability for a single mutant is\cite{antal2006fixation,traulsen2009stochastic}

\begin{equation}
    \Phi_{1,N} = \dfrac{1}{1 + \sum\limits_{k=1}^{N-1} \prod\limits_{j=1}^{k} \dfrac{g_i}{f_i}}.
    \label{efp2}
\end{equation}
The same recursive equation applies for the absorption time $a_{i}$ and fixation time $t_{i,N}$ as follows:
\begin{align}
        \begin{cases}
        a_i=p_{i \longrightarrow  i-1}a_{i-1}+(1-p_{i \longrightarrow  i-1}-p_{i \longrightarrow  i+1})a_{i}+p_{i \longrightarrow  i+1}a_{i+1}, \\
        \Phi_{i,N}t_{i,N}=p_{i \longrightarrow  i-1}\Phi_{i-1,N} (t_{i-1,N}+1)+(1-p_{i \longrightarrow  i-1}-p_{i \longrightarrow  i+1})\Phi_{i,N} (t_{i,N}+1)\\
        \hspace{40pt}+p_{i \longrightarrow  i+1}\Phi_{i+1,N} (t_{i+1,N}+1).  
    \end{cases}
    \label{fixt1}
\end{align}

With the initial conditions $a_0=t_{0,N}=0$ and $a_N=t_{N,N}=1$ we have\cite{antal2006fixation,traulsen2009stochastic}: 

\begin{align}
    \begin{cases}
        a_{1} = \Phi_{1,N} \sum\limits_{k=1}^{N-1} \sum\limits_{l=1}^{k} \frac{1}{P_{l \rightarrow l+1}} \prod\limits_{j=l+1}^{k} \dfrac{g_j}{f_j} ,\\[10pt]
        t_{1,N} = \sum\limits_{k=1}^{N-1} \sum\limits_{l=1}^{k}  \frac{\Phi_{l,N}}{P_{l \rightarrow l+1}} \prod\limits_{j=l+1}^{k} \dfrac{g_j}{f_j} .
    \end{cases}
    \label{efixt1}
\end{align}

\section{General Framework to calculate the absorbing probability and time for Markov chains}\label{farmwork}
To evaluate the absorption probabilities and time for Markov chains with $n=t+a$ states ($t$ transition state and $a$ absorbing states), a widely employed technique in numerical analysis, as outlined in \cite{grinstead1997introduction,hindersin2014counterintuitive}, can be employed. The transition matrix $P_{n*n}$ can be reexpressed in a canonical form as follows:
\begin{equation}
P= 
\begin{pmatrix}
  Q & R\\ 
  0 & I
\end{pmatrix},
\label{rep}
\end{equation} 
where $Q_{t*t}$ denotes the probabilities of transitions between the transient states, $R_{t*a}$ signifies the probabilities of transitions from the transient states to the absorbing states and $I_{a*a}$ represents an identity matrix.  The Moran process in a well-mixed population is related to a Markov chain with $N+1$ states, including two absorbing states at $0$ and $N$. 
Based on the canonical transition matrix, we can construct a new matrix denoted by $F=\sum_{n=0}^{\infty}Q^n=(I-Q)^{-1}$ and called a Fundamental matrix. The element $F_{i,j}$ represents the expected number of steps that the process which started in state $i$ spends in state $j$ before eventually reaching one of the absorbing states (the expected {\it sojourn time}).
Based on this matrix, we can derive another matrix:
\begin{equation}
\Phi=FR,
\label{4}
\end{equation}
where $\Phi_{i,j}$ represents the probability that the process becomes absorbed in absorbing state $j$, given that it starts in state $i$. For a Moran process in a well-mixed population, the matrix $\Phi$ is composed of two columns, representing extinction and fixation probabilities. 

The absorption time starting from state $i$
can be obtained by summing the $i_{th}$ row of the fundamental matrix $F$ and is given by the following expression:
\begin{equation}
a_i=\sum_{j=1}^{N-1}F_{i,j}
\label{maruntime}
\end{equation}

To determine the fixation time, we employ a method described in previous studies\cite{ewens1973conditional,hindersin2014counterintuitive,altrock2012mechanics}. If $\Phi_{j,N}$
represents the fixation probability starting from $j$ mutants, then the  fixation time starting from state $i$ can
be calculated as follows:
\begin{equation}
t_{i,N}=\sum_{j=1}^{N-1}\dfrac{\phi_{j,N}}{\phi_{i,N}} F_{i,j}
\label{6}
\end{equation}
By identifying the Markov chain and associated transition matrix for various models, we can utilize this numerical method to determine fixation probability and times.
\section{Model and methods} \label{model}

A random process should have a starting point. In our model, this is the time that a single mutant appears in the population, labelled time $t=0$. In the above we assume that the process is homogeneous, so that transition probabilities depend only upon the current state and not on the specific time that the state is reached. Thus for example in equation \ref{efp1} the transition and fixation probabilities are independent of the specific time point, and so are not functions of time $t$. In the model developed below we will similarly consider the process homogeneous, but develop the concept of the state of the process to include not only the population size at present, but also at the most recent times, to incorporate time delays. This still leaves the question of how to deal with times too close to the start of the process for the full required history to exist. We explain how we do this, by selecting a specific initial state, in Section \ref{sec5a}.

Now we introduce a time delay in the Bd process for a well-mixed population. In the standard Bd process, the fitness of strategies at the current (discrete) time $t$ is important for the birth of new offspring. However, an individual's fitness can depend on conditions in the past. Thus, we incorporate a time delay $\tau$, so that the fitness of strategies comes from time $t-\tau$, considering the same time delay for all strategies. Thus, both mutant and resident strategies receive payoffs from $\tau$ steps before. Since the Bd process proceeds in discrete time steps, $\tau$ takes integer values in the $[0,\infty)$ range. The case $\tau=0$ corresponds to the standard Bd process described previously.

To study this process, we model it as a Markov chain and use the previous numerical approach. To fully characterize each Markov chain state we need to know both the number of mutants at the current time $t$ and the historical ones from time $t-\tau$ to the present. This allows us to define the state space describing the full process dynamics. Specifically, each state has  $\tau+1$ indices, with $\tau$ indices representing the number of mutants at each of the $\tau$ previous time steps, and one index for the current number at time $t$. When $\tau=0$, the state reduces to a single index, as originally discussed. With the general state form $\{i_{\tau},...,i_1,i_0\}$, the number of mutants can only change by one per time step. This imposes $|i_n-i_{n-1}|\leq 1$ between adjacent indices. As an example, for $\tau=1$ each state is characterized by $\{i_1,i_0\}$, with $i_0=i_1-1,i_1,i_1+1$. Within this Markov chain, states characterized by $i_0=0, N$ serve as absorbing states, signifying mutant extinction and fixation. 

\begin{figure}[H]
{\includegraphics[scale=0.33]{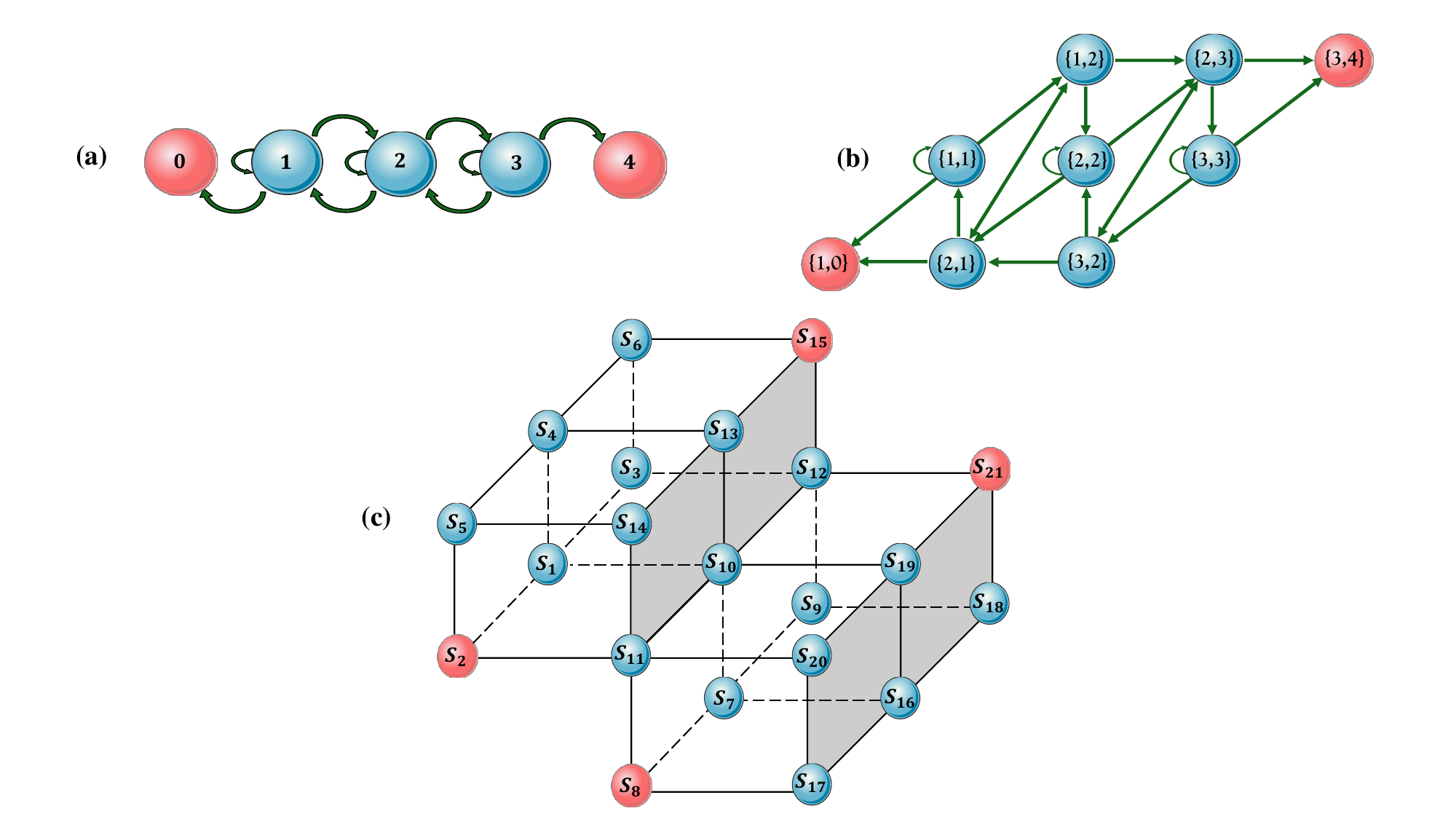}}
\centering
\caption{Three Markov chains for $\tau=0,1,2$ with population size $N=4$. The Markov chain with time delays $\tau$ has $\tau+1$ dimensions. To generate state spaces for a Markov chain with $\tau+1$ dimensions, we must erase the absorbing states from the $\tau$-dimensional Markov chain and add three possible new indexes for each state $\{i_\tau,....,i_0\}$ like $\{i_\tau,....,i_0,i_{new}\}$  including $i_{new}=i_0-1,i_0,i_0+1$(See {\textbf{Algorithm 1}}). For instance, a population with size $N=4$ have (a):$S_4^0=\big\{\{0\},\{1\},\{2\},\{3\},\{4\}\big\}$; (b): $S_4^1=\big\{\{1,0\},\{1,1\},\{1,2\},\{2,1\},\{2,2\},\{2,3\},\{3,2\},\{3,3\},\{3,4\}\big\}$; and (c): $S_4^2=\big\{s_1=\{1,1,1\},s_2=\{1,1,0\},s_3=\{1,1,2\},s_4=\{1,2,2\},s_5=\{1,2,1\},s_6=\{1,2,3\},s_7=\{2,1,1\},s_8=\{2,1,0\},s_9=\{2,1,2\},s_{10}=\{2,2,2\},s_{11}=\{2,2,1\},s_{12}=\{2,2,3\},s_{13}=\{2,3,3\},s_{14}=\{2,3,2\},s_{15}=\{2,3,4\},s_{16}=\{3,2,2\},s_{17}=\{3,2,1\},s_{18}=\{3,2,3\},s_{19}=\{3,3,3\},s_{20}=\{3,3,2\},s_{21}=\{3,3,4\}\big\}$. The transition and absorbing states are represented by blue and red circles, respectively. When $\tau>1$, there are more absorbing states with $i_0=N, 0$ indicating mutant fixation and extinction respectively.}
\label{fmar}
\end{figure}

Deriving the whole state space of the Markov chain $S_N^\tau={s_i}$ for a time delay of $\tau$  in a population of size $N$ can be accomplished through the subsequent procedure.
\begin{equation}
\begin{split}
&S_N^0=\big\{\{i_0\}\big\},\hspace{9pt} i_0\in \{0,1,2,...,N\}, \hspace{264pt}  \\
&S_N^1=\big\{\{i_1,i_0\}\big\},\hspace{9pt} i_1\in S_N^0, \hspace{9pt} i_1 \neq \{0,N\}, \hspace{9pt} i_0 \in \{i_1+1,i_1,i_1-1\}, \hspace{124pt} \\
&S_N^2=\big\{\{i_2,i_1,i_0\}\big\},\hspace{9pt} \{i_2,i_1\}\in S_N^1, \hspace{9pt} \{i_2,i_1\} \neq \big\{\{1,0\},\{N-1,N\}\big\}, \hspace{9pt} i_0 \in \{i_1+1,i_1,i_1-1\}, \hspace{5pt}\\
&...\hspace{200pt} \\
&...\hspace{200pt}\\
&S_N^\tau=\big\{\{i_\tau,i_{\tau-1},...,i_1,i_0\}\big\},\hspace{9pt} \{i_\tau,i_{\tau-1},...,i_1\}\in S_N^{\tau-1}, \hspace{9pt} \{i_\tau,i_{\tau-1},...,i_1\} \neq \big\{\text{Absorbing States}\big\}, \\ & \hspace{235pt}i_0 \in \{i_1+1,i_1,i_1-1\}.
\end{split}
\label{samplestate}
\end{equation}
The previous procedure allows us to find the state space of $S_N^\tau$ based on $S_N^{\tau-1}$. As shown in Fig.\ref{fmar}, state space $S_N^\tau$ can be represented as a $\tau+1$ dimensional Markov chain. Simply we can obtain states of $S_N^\tau$ from $S_N^{\tau-1}$. We identify and delete the absorbing states of $S_N^{\tau-1}$ where $i_0=0$ or $N$, representing extinction or fixation. The remaining states of $S_N^{\tau-1}$ become the historical component of each state in $S_N^\tau$ ($i_j$ in $S_N^{\tau-1}$ corresponding to $i_{j+1}$ in $S_N^\tau$). Since the number of mutants can change by at most one, $i_0$ in $S_N^\tau$ takes values $i_1-1$, $i_1$, or $i_1+1$ ($i_1$ in $S_N^\tau$  is $i_0$ in $S_N^{\tau-1}$). The process excludes the absorbing states corresponding to $S_N^{\tau-1}$ and adds one more dimension (one more index to each state satisfying previous conditions) to find the Markov chain of $S_N^\tau$. 
{\textbf{Algorithm 1}} can be used to find the state space for a given population size and time delay.

\begin{figure}
{\includegraphics[scale=0.08]{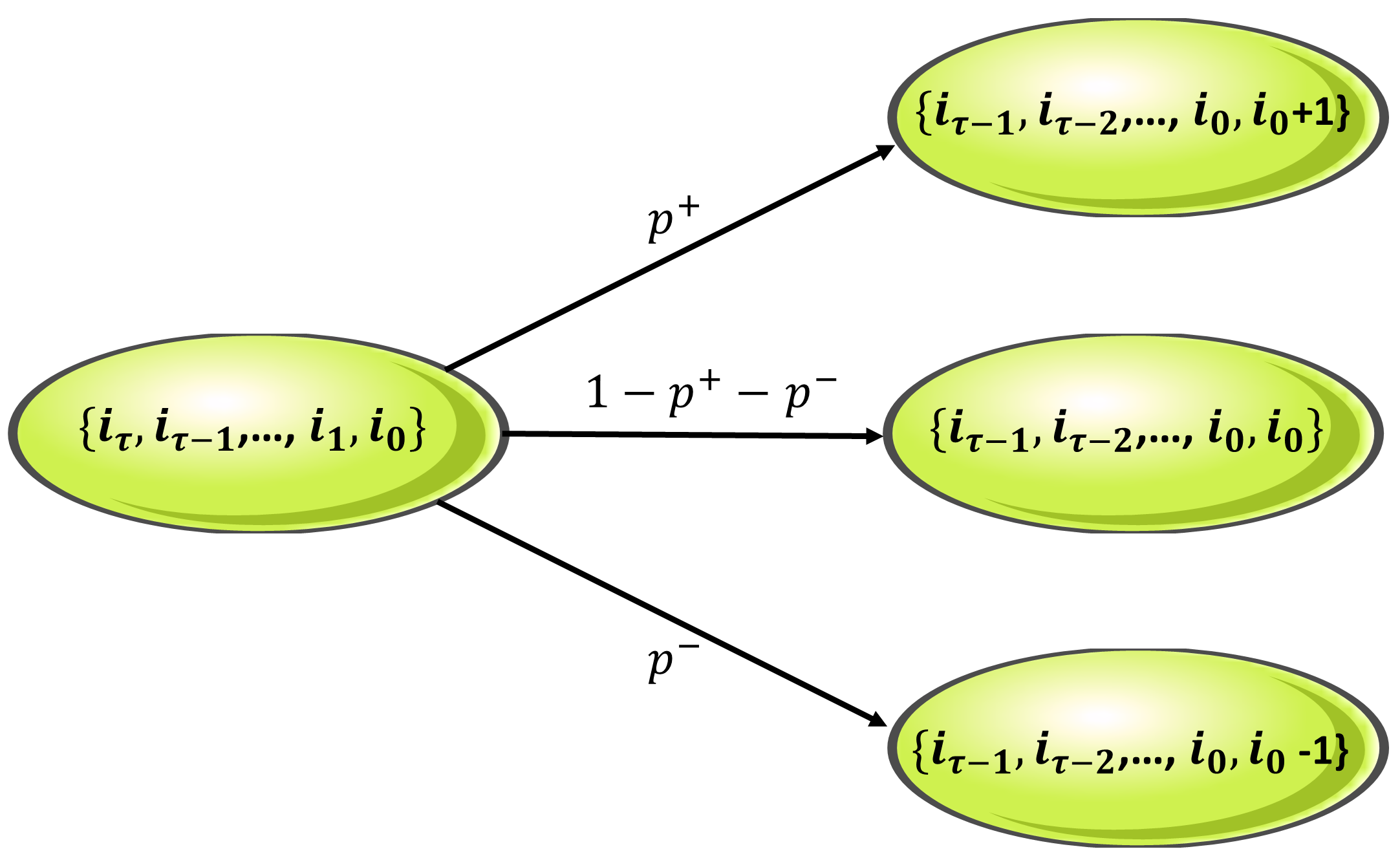}}
\centering
\caption{There are three possible transitions from state $\{i_\tau,i_{\tau-1},...,i_1,i_0\}$ in which the number of mutants in the population represented by $i_0$ can rise by one, drop by one, or remain unchanged. in each transition, the history will transform in such a way that $\{i_\tau,i_{\tau-1},...,i_1,i_0\} \longrightarrow \{i_{\tau-1},i_{\tau-2},...,i_1,i_0,i_{new}\}, i_{new}=i_0+1,i_0,i_0-1 $.  }
\label{ftra}
\end{figure}

The next step is to find the transition matrix for the Markov chain with state space $S_N^\tau$. Since we are considering the Bd process, time delays can affect the Birth part where fitness for reproduction comes from the past. If we denote the state as $\{i_\tau, i_{\tau-1},..., i_1, i_0\}$, there are three possible transitions as shown in Fig. \ref{ftra}. In each generation of the process the number of mutants $i_0$ can either increase by one, decrease by one, or remain unchanged. In each transition, the history changes such that the new $i_j$ becomes $i_{j-1}$ from the previous state. To find the transition probability, the fitness for the birth part comes from $i_\tau$ which shows the number of mutants $\tau$ steps ago. However, for the death part, the number of current mutants $i_0$ is important. Considering these details, the transition probabilities are:
\begin{equation}
\begin{split}
p^+=p_{\{i_\tau,i_{\tau-1},...,i_1,i_0\} \longrightarrow \{i_{\tau-1},i_{\tau-2},...,i_0,i_0+1\}}=\dfrac{i_0f_{i_\tau}}{i_0f_{i_\tau}+(N-i_0)g_{i_\tau}}\dfrac{N-i_o}{N-1}\\
p^-=p_{\{i_\tau,i_{\tau-1},...,i_1,i_0\} \longrightarrow \{i_{\tau-1},i_{\tau-2},...,i_0,i_0-1\}}=\dfrac{(N-i_0)g_{i_\tau}}{i_0f_{i_\tau}+(N-i_0)g_{i_\tau}}\dfrac{i_o}{N-1}\\
p_{\{i_\tau,i_{\tau-1},...,i_1,i_0\} \longrightarrow \{i_{\tau-1},i_{\tau-2},...,i_0,i_0\}}=1-p^+-p^-\hspace{96pt}
\end{split}
\label{etransitionprobability}
\end{equation}
where $f_{i_\tau}$ and $g_{i_\tau}$ are given in \eqref{efitnessfunction}. \textbf{Algorithm 2} provides a useful way to determine the transition matrix $P^\tau$ for the time delay $\tau$. Once $P^\tau$ is obtained, we can identify all absorbing states where $i_0=0,N$ and put the matrix into canonical form $F^\tau$ by finding $Q^\tau$ and $R^\tau$. This then allows the numerical approach from the previous section to be applied to calculate the fixation probability and time. As mentioned before, there can be additional fixation states (depending on $\tau$) with $i_0=N$. Since the sum of each row of the absorbing probability matrix, $\phi^\tau=F^\tau R^\tau$ is 1 (the sum of absorption probabilities starting from one state), we can find the fixation probability by summing the probabilities leading to fixation with $i_0=N$. By dividing the state space $S_N^\tau$ into two parts $S_T^\tau \subset S_N^\tau, i_0 \neq 0, N$ for transient states and $S_A^\tau \subset S_N^\tau, i_0 = 0, N$ for absorbing states, we can define the fixation probability $\phi_{s_m,N}^\tau$ and time $t_{s_m,N}^\tau$ starting from state $s_m$ as:

\begin{equation}
\Phi^\tau=F^\tau R^\tau, \hspace{20pt} \phi_{s_m,N}^\tau=\sum_{s_j=\{i_\tau,...,i_0\},i_0=N}^{S_A^\tau}\Phi_{s_m,s_j}^\tau,
\label{9}
\end{equation}
\begin{equation}
t_{s_m,N}^\tau=\sum_{s_j=\{i_\tau,...,i_0\}}^{S_T^\tau}\dfrac{\phi_{s_j,N}^\tau}{\phi_{s_m,N}^\tau} F_{s_m,s_j}^\tau,
\label{10}
\end{equation}

\begin{table}[ht]
  \begin{minipage}[b]{0.45\linewidth}
    \centering
   \resizebox{0.9\columnwidth}{!}{
    \begin{tabular}{l}
      \hline
      \textbf{Algorithm 1. Find the state space} \\
      \hline
1: \textbf{Input:} Size N and time delay $\tau$ \\

2: state-space $\leftarrow$ empty-array(N+1)\\

3: \textbf{for} $l$ \textbf{in} $[0,1,2,...,N]$ \textbf{do} \\

4: \hspace{10pt}  state-space($l$) $\leftarrow$ $l$  \\

5: \textbf{end for}\\

6: \textbf{for} $t$ \textbf{in} $[0,1,2,...,\tau-1]$ \textbf{do} \\

7: \hspace{10pt}  \textbf{for} state \textbf{in} state-space \textbf{do} \\

8: \hspace{25pt}  \textbf{if} $i_0$=0 \textbf{or} $i_0$=N \textbf{then}\\

9: \hspace{38pt} \textbf{remove} state \textbf{from} sample-state\\

10: \hspace{22pt}  \textbf{end if}\\

11: \hspace{10pt}  \textbf{end for} \\

12: \hspace{11pt} new-state-space $\leftarrow$ empty-array(size(state-space)$\times$ 3,t+1)\\

13:\hspace{10pt} \textbf{for} state \textbf{in} state-space \textbf{do}\\

14: \hspace{22pt}  \textbf{add} \{state,$i_0$-1\} \textbf{and} \{state,$i_0$\} \textbf{and} \{state,$i_0$+1\}\\

\hspace{41pt} \textbf{to} new-state-space\\

15:\hspace{10pt} \textbf{end for}\\

16:\hspace{11pt} state-space$ \leftarrow$ new-state-space\\

17: \textbf{end for} \\

17: \textbf{return} state-space \\
\hline
    \end{tabular}}
  \end{minipage}
  \hfill
  \begin{minipage}[b]{0.45\linewidth}
    \centering
    \resizebox{\columnwidth}{!}{
    \begin{tabular}{l}
      \hline
      \textbf{Algorithm 2. Find the transition matrix} \\
      \hline
1: \textbf{Input:} state-space \\

2: transition-matrix $\leftarrow$ zero-array(size(state-space) $\times$ size(state-space))\\

3: \textbf{for} state \textbf{in} state-space \textbf{do}\\

4:\hspace{10pt} \textbf{if} state=absorbing-state \textbf{then}\\

5: \hspace{20pt} transition-matrix(state, state)=1\\

6:\hspace{10pt} \textbf{else}:\\

7: \hspace{20pt} \textbf{for} second-state  \textbf{in} state-space \textbf{do}  \\

8: \hspace{30pt} \textbf{if} $\{i_\tau,i_{\tau-1},...,i_1\}$ $\in$ second-state= $\{i_{\tau-1},...,i_1,i_0\}$ $\in$ state \textbf{then}\\

9: \hspace{45pt} \textbf{if} ($i_0\in \text{second-state}$ - $i_0\in \text{state}$)=1 \textbf{then}\\

10: \hspace{50pt} transition-matrix(state, second-state)=$p^+$\\


11: \hspace{40pt} \textbf{else if} ($i_0\in \text{second-state}$ - $i_0\in \text{state}$)=-1 \textbf{then}\\

12: \hspace{50pt} transition-matrix(state, second-state)=$p^-$\\


13: \hspace{40pt} \textbf{else}\\

14: \hspace{50pt} transition-matrix(state, second-state)=$1-p^--p^+$\\

15: \hspace{40pt} \textbf{end if} \\

16: \hspace{30pt} \textbf{end if} \\

17: \hspace{20pt} \textbf{end for} \\

18: \hspace{8pt} \textbf{end if} \\

18: \textbf{end for} \\

19: \textbf{return} transition-matrix \\
\hline
    \end{tabular}}
  \end{minipage}
\end{table}
\section{Results}\label{results}
\begin{figure}[H]
{\includegraphics[scale=0.53]{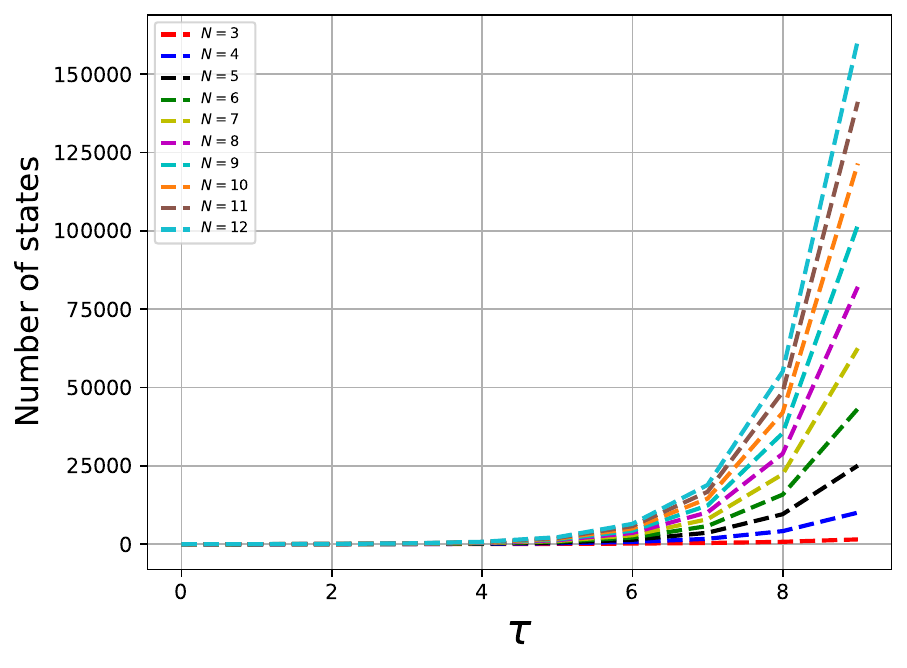}}
\centering
\caption{The number of states as a function of $N$ and $\tau$.}
\label{fig22}
\end{figure}

In this section, we examine how time delay impacts the fixation of mutants. Before presenting the main results, it is instructive to demonstrate some limitations of our numerical method. As Fig. \ref{fig22} shows, the state space size grows with both $N$ and $\tau$. For large $\tau$ and $N$ values, the number of states becomes enormous, necessitating more memory to store the data. Furthermore, calculating the inverse matrix has higher computational complexity as the matrix size increases. For instance, with $N=5$ and $\tau=6$, there are 1,398 states, yielding a  matrix with size of $1,954,404$. This matrix would require approximately $15.6$ MB of memory ($1,954,404 \times 8$  bytes). However, when $N$ grows to $12$ and $\tau$ to $9$, the state space has $160,875$ states, requiring around $207$ GB of memory—a prohibitive amount. Clearly, for larger populations and time delays, our numerical method has limitations, and computer simulations become preferable.

\subsection{Fixation probability and time}\label{sec5a}

In the following, we investigate the effect of time delays on the fixation probability and time in some well-known games; the Stag-Hunt, Snowdrift, and Prisoner's Dilemma.  When studying the fixation of one mutant in a population of residents, there may be multiple states with $i_0=1$ that differ in their historical components for a given time delay $\tau$. For consistency, we will use the initial state $s_i=\{1,1,...,1\}$ for all our results; this is consistent with starting with a single mutant and that for times within $\tau$ of the start of our process, the payoffs used to calculate the fitness are assumed to take the value at this starting point. We compare the results of our numerical method (as much as possible) to those from computer simulations. For the simulations, we conduct $10^7$ different Monte Carlo simulations for each set of fixed parameters. The fixation probability is calculated as the ratio of fixed processes to total simulations. We determine the average fixation time only using processes that eventually undergo fixation.

\subsubsection{Stag-Hunt game}

The Stag-Hunt game is one of the earliest and most analyzed coordination games\cite{skyrms2004stag}. This game is defined by a payoff matrix where the payoffs satisfy the relation $a > c \geq d > b$. The key feature of the Stag-Hunt is that players are incentivized to coordinate on the same strategy, represented by the payoffs $a$ and $d$ along the leading diagonal of the payoff matrix. However, there is also a risk, captured by the off-diagonal payoffs $b$ and $c$, if the players fail to match strategies. To minimize the number of parameters, we represented the payoff matrix for this game as follows:
\begin{center}
\begin{tabular}{ c c c c }
 &      & A & B \\ 
    & A & $\alpha$ & $\beta$ \\
$U_1$ = & & &     \\  
    & B & $\alpha\beta$ & $\alpha\beta$    
\end{tabular}
\end{center}

where $\alpha>1$ and $0<\beta<1$.

\begin{figure}
{\includegraphics[scale=0.2]{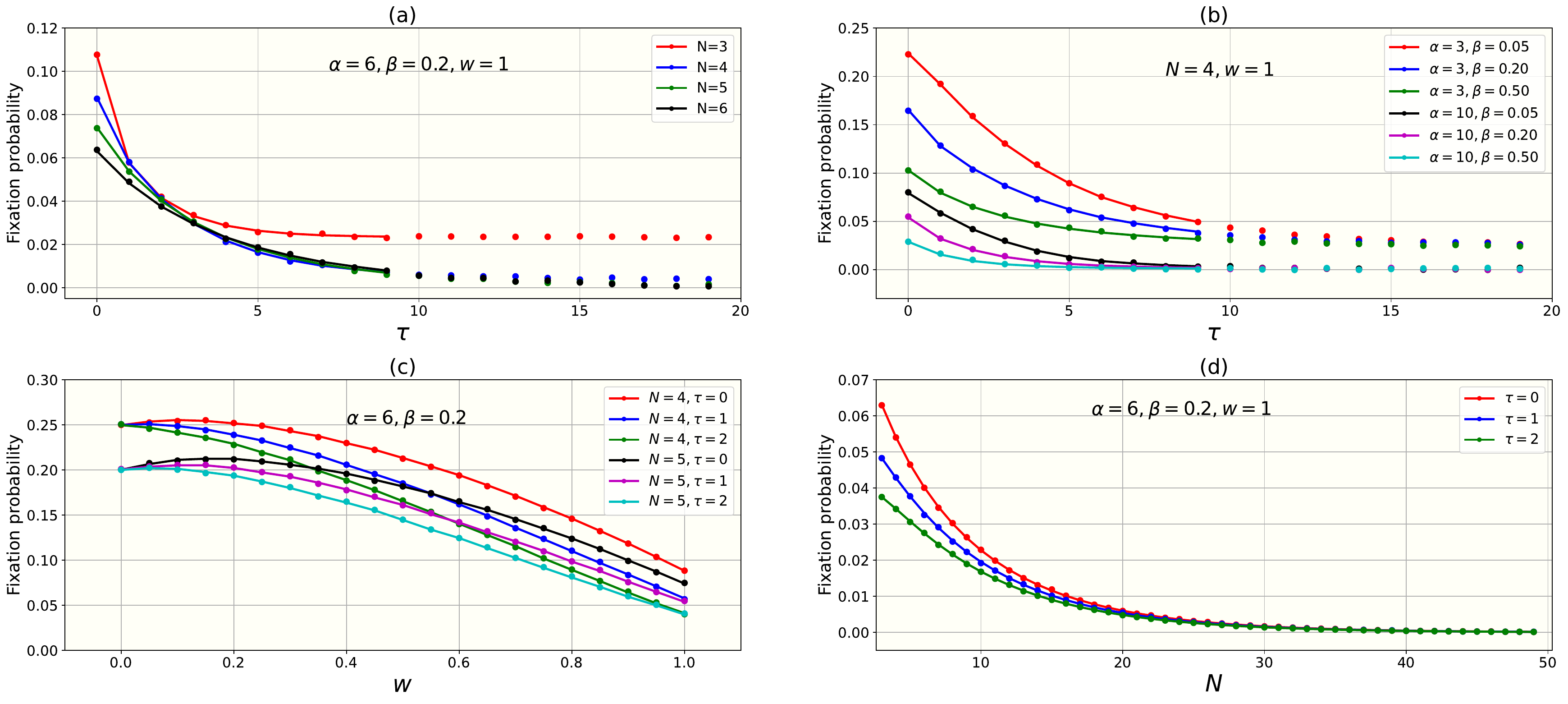}}
\centering
\caption{The effect of time delays on fixation probabilities in the Stag-Hunt game ($U_1$) according to numerical solutions (lines) and computer simulations(dots). Panel (a) shows the influence of time delays for various network sizes with $\alpha=6$, $\beta=0.2$, and $w=1$. Panel (b) illustrates the effect of time delays for different payoff matrix parameters with $N=4$ and $w=1$. Panel (c) depicts how selection intensity affects outcomes for $N=4,5$ and $\tau=0,1,2$ with $\alpha=6$, $\beta=0.2$. Lastly, panel (d) demonstrates the impact of network size for $\tau=0,1,2$ with $\alpha=6$, $\beta=0.2$ and $w=1$.}
\label{figstapr}
\end{figure}

\begin{figure}
{\includegraphics[scale=0.2]{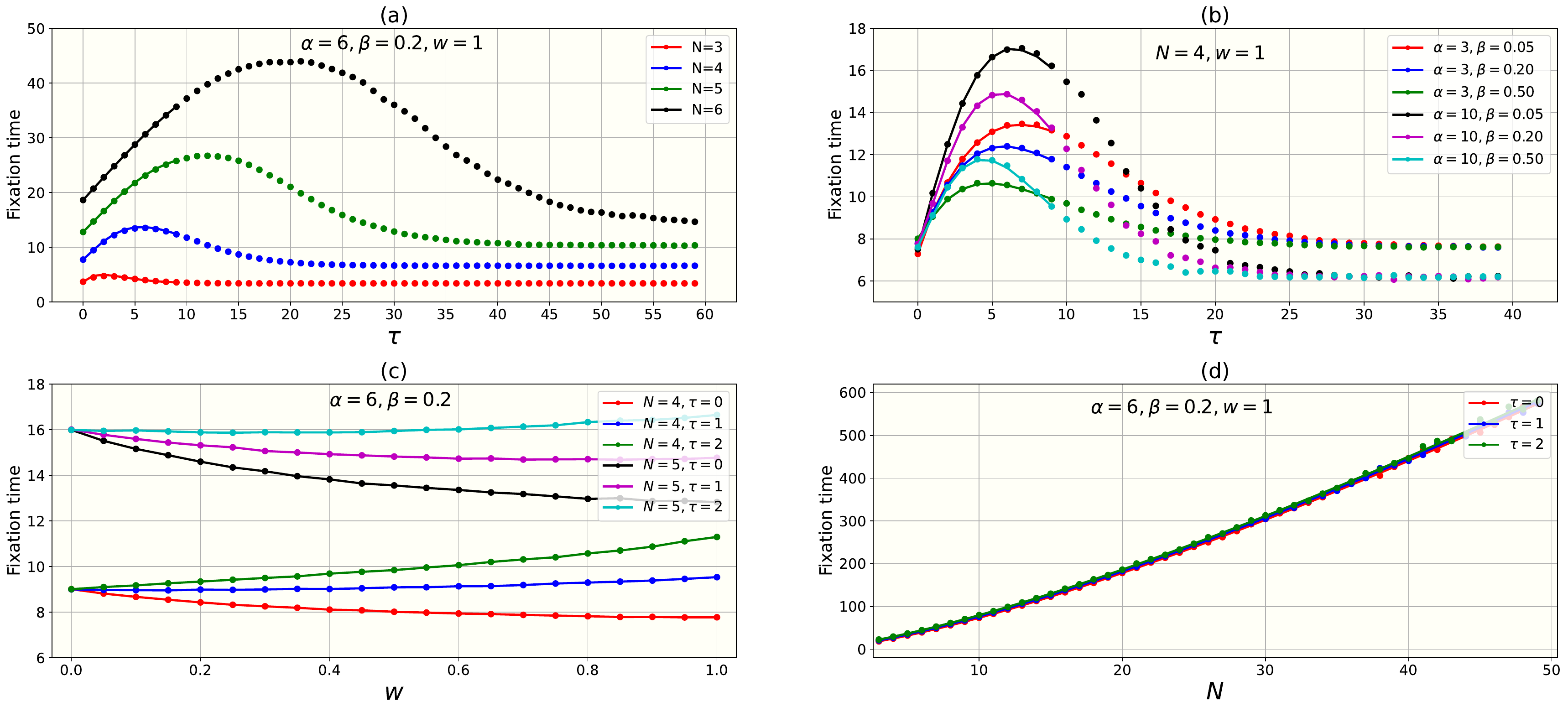}}
\centering
\caption{The effect of time delays on fixation time in the Stag-Hunt game ($U_1$) according to numerical solutions (lines) and computer simulations(dots). Panel (a) shows the influence of the time delays for various network sizes with $\alpha=6$, $\beta=0.2$, and $w=1$. Panel (b) illustrates the effect of time delays for different payoff matrix parameters with $N=4$ and $w=1$. Panel (c) depicts how selection intensity affects outcomes for $N=4,5$ and $\tau=0,1,2$ with $\alpha=6$, $\beta=0.2$. Lastly, panel (d) demonstrates the impact of network size for $\tau=0,1,2$ with $\alpha=6$, $\beta=0.2$, and $w=1$.}
\label{figstati}
\end{figure}

We first analyze how time delay impacts the fixation probability when there is one mutant with strategy A (the ``stag'' strategy) in a population of residents using strategy B. Fig. \ref{figstapr} shows how varying the payoff matrix elements, population size, and selection intensity affects the fixation probability in the presence of time delays. The lines represent numerical solutions while the dots show computer simulations. As noted previously, our numerical method is limited computationally, restricting the population size and time delay values we could examine. Thus for larger $N$ and $\tau$, only simulation results are presented. Regardless of the network size and payoff element values, longer time delays consistently yield lower fixation probabilities.  The results demonstrate that time delays reduce the ability of a mutant to take over a population compared to the undelayed cases. 

Fig. \ref{figstapr} (a) and (b) demonstrate that as $\tau$ increases, the fixation probabilities appear to converge to a constant value. This suggests that especially for a population of fixed size, the fixation probability at large $\tau$ depends only on $\alpha$, not $\beta$. Our model indicates that as $\tau$ approaches infinity, fitness remains unchanged throughout the fixation process. This is because fitness originates from the starting state with one mutant and persists across all steps until fixation. Thus, we have a Moran process with constant fitness, where the fixation probability follows Eq \eqref{efp2}. With $w=1$ and the constant fitness defined as $r$:
\begin{equation*}
\begin{split}
\tau \rightarrow \infty
\begin{cases}
    \pi_A(i)=\pi_A(1)=\dfrac{(N-1)\beta}{N-1} \rightarrow f_i=\beta\\
    \pi_B(i)=\pi_B(1)=\dfrac{(N-1)\alpha\beta}{N-1} \rightarrow  g_i=\alpha\beta \hspace{15pt}
\end{cases}
r=\dfrac{f_i}{g_i}=\dfrac{1}{\alpha}.
\end{split}
\label{eqcofi1}
\end{equation*}
Therefore the fixation probability with constant fitness $r$ is:
\begin{equation}
    \Phi_{1,N} = \dfrac{1}{1 + \sum\limits_{k=1}^{N-1} \prod\limits_{j=1}^{k} \dfrac{1}{r}}.=\dfrac{1-\dfrac{1}{r}}{1-{(\dfrac{1}{r})^N}}=\dfrac{1-\alpha}{1-{(\alpha)}^N}.
    \label{eqcofi2}
\end{equation}

For sufficiently large values of $\tau$, the fixation probability depends only on $\alpha$ and $N$. When $\alpha=6$, the fixation probability is $\frac{1-6}{1-6^N}$. For $N=3,4,5,6$, this gives fixation probabilities of $0.02325$, $0.00386$, $0.00064$, and $0.00010$ respectively, matching Fig. \ref{figstapr} panel (a). With a constant population size $N=4$ in panel (b), increasing $\tau$ causes the fixation probability to converge to $\frac{1-\alpha}{1-\alpha^4}$. Thus the fixation probability is only affected by $\alpha$, which is the ratio of fitness between the resident and mutant types at the initial point when there is just one mutant individual present in the population. In the large $\tau$ limit, the fixation probabilities for $\alpha=3,10$ are $0.025$ and $0.0009$. 
The influence of the intensity of selection $w$ is shown in panels (c) for a fixed payoff matrix with $\alpha=6$ and $\beta=0.2$. When $w=0$, the payoff has no effect on fitness, resulting in natural selection with a fixation probability of $1/N$. As $w$ increases, the role of the payoff and the time delay becomes more apparent. Notably, the fixation probability initially rises with increasing $w$ before later decreasing. Panel (d) demonstrates the effect of the population size for three time delay values. For sufficiently large populations, the impact of the time delay on the fixation probability is negligible, with fixation probabilities almost zero in each case. However, for small populations, the time delay has a considerable effect on the fixation probability.

Fig. \ref{figstati} shows how time delays affect the fixation time in this game. Panel (a) demonstrates the impact of $\tau$ for various population sizes. Initially, increasing the time delay at a constant population size causes the fixation time to rise. However, after a critical $\tau$ value, the time delay starts to reduce fixation time until it levels off at a fixed value for very large $\tau$. As discussed previously, a sufficiently long time delay produces a Moran Bd process with fixed fitness derived from the initial state with $1$ mutant. Therefore, for very large $\tau$, the fixation time can be calculated using Eq \eqref{efixt1}. For the Stag-Hunt game defined by $U_1$, the ratio of $\frac{f_i}{g_i}$ in \eqref{efixt1} is $r=\frac{1}{\alpha}$.

The influence of the time delay on the fixation time in a population with constant size for different parameters of the payoff matrix is illustrated in panel (b) of Fig.\ref{figstati}. As the time delay increases, the fixation time rises to a maximum before decreasing. Ultimately, all fixation times converge to a constant value for the same $\alpha$. Panels (c) and (d) show how the intensity of selection $w$ and population size impact the fixation time for certain time delays. As the game becomes more influential in determining fitness and selection intensity increases, the fixation time varies with the time delay. Additionally, for a fixed time delay, increasing the population size leads to greater fixation times, with network size having a larger effect than the time delay.

\subsubsection{Snowdrift game}
Generally, a Snowdrift game is characterized by a payoff matrix for which $c > a > b > d$ \cite{broom2022game,doebeli2005models}. This setting of the payoff matrix promotes cooperation (playing strategy A) by an individual when they encounter a defecting opponent (playing strategy B). The game describes two people who are stuck in a snowdrift and both want to get out of it but are not eager to share the cost of clearing the snow. In literature, the Snowdrift game can be parameterized as follows:
\begin{figure}
{\includegraphics[scale=0.2]{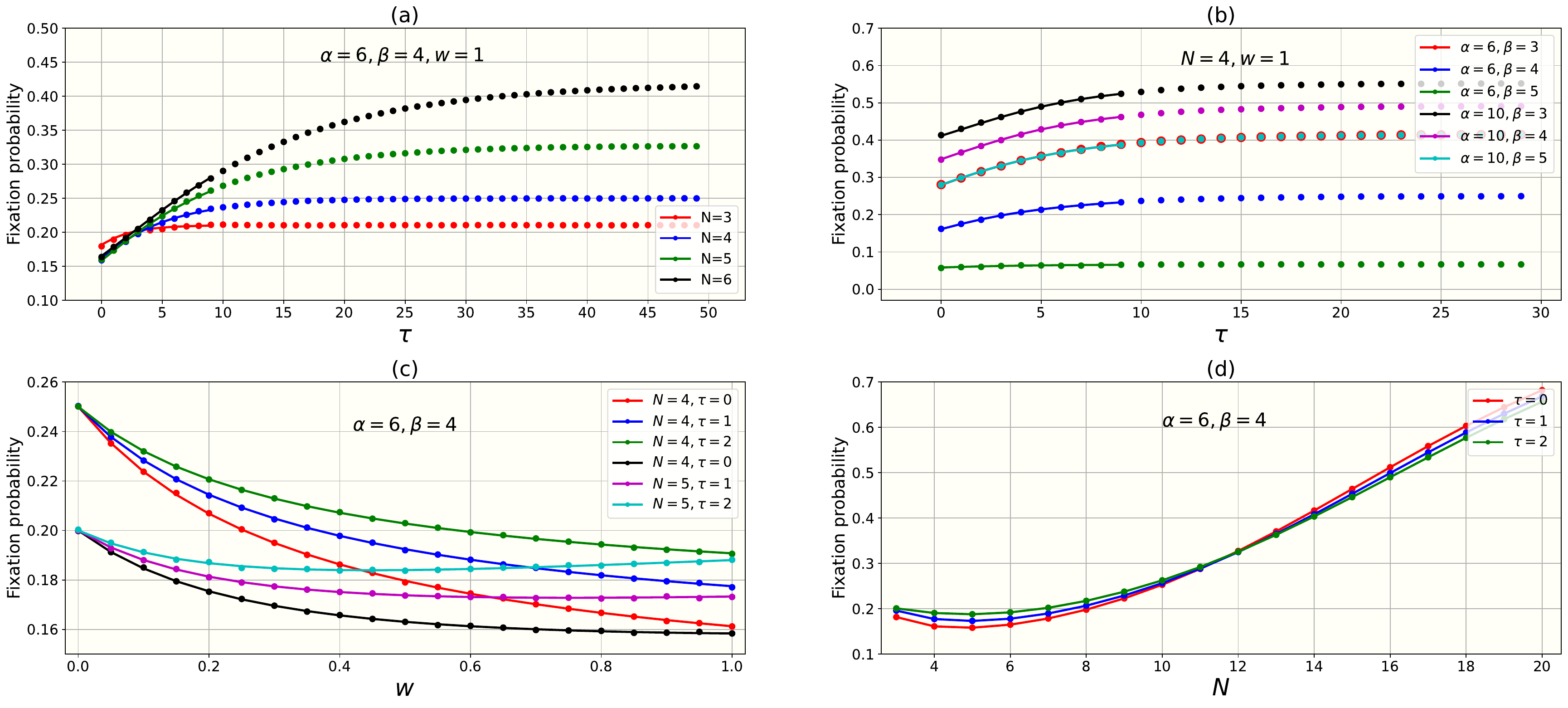}}
\centering
\caption{The effect of time delays on the fixation probabilities in the Snowdrift game ($U_2$) according to numerical solutions (lines) and computer simulations(dots). Panel (a) shows the influence of time delays for various network sizes with $\alpha=6$, $\beta=4$, and $w=1$. Panel (b) illustrates the effect of time delays for different payoff matrix parameters with $N=4$ and $w=1$. Panel (c) depicts how selection intensity affects outcomes for $N=4,5$ and $\tau=0,1,2$ with $\alpha=6$, $\beta=4$. Lastly, panel (d) demonstrates the impact of network size for $\tau=0,1,2$ with $\alpha=6$, $\beta=4$, and $w=1$.}
\label{figsdpr}
\end{figure}

\begin{figure}
{\includegraphics[scale=0.2]{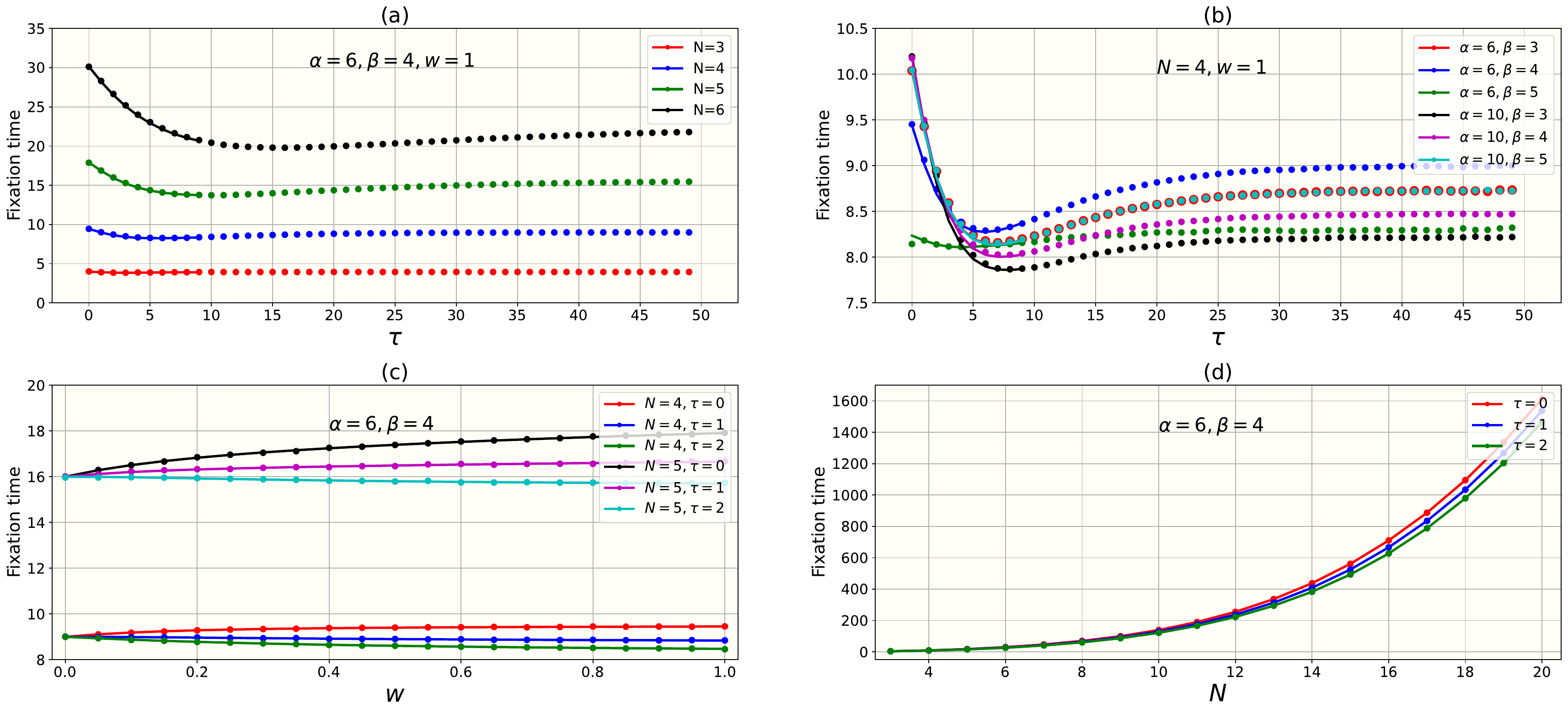}}
\centering
\caption{The effect of time delay on fixation time in the Snowdrift game ($U_2$) according to numerical solutions (lines) and computer simulations(dots). Panel (a) shows the influence of time delay for various network sizes with $\alpha=6$, $\beta=4$, and $w=1$. Panel (b) illustrates the effect of time delay for different payoff matrix parameters with $N=4$ and $w=1$. Panel (c) depicts how selection intensity affects outcomes for $N=4,5$ and $\tau=0,1,2$ with $\alpha=6$, $\beta=4$. Lastly, panel (d) demonstrates the impact of network size for $\tau=0,1,2$ with $\alpha=6$, $\beta=4$, and $w=1$.}
\label{figsdti}
\end{figure}

\begin{center}
\begin{tabular}{ c c c c }
 &      & A & B \\ 
    & A & $\alpha-\frac{\beta}{2}$ & $\alpha-\beta$ \\
$U_2$ = & & &     \\  
    & B & $\alpha$ & 0    
\end{tabular}
\end{center}
where $\alpha>\beta$.

The time delay impacts the fixation probability in the Snowdrift game, as shown in Fig. \ref{figsdpr}. The top panels illustrate how increasing $\tau$ affects the fixation probability for various payoff parameters and population sizes. Unlike the Stag-Hunt game, a longer time delay boosts fixation probability in the Snowdrift game. Notably in panel (a), a small $\tau$ leads to a lower value in larger populations, but as $\tau$ grows, larger populations exhibit a greater value. In both panels, fixation probabilities plateau at constant values dependent on population size and game parameters when $\tau$ is sufficiently high. Similar calculations to the Stag-Hunt game reveal that:

\begin{equation*}
\begin{split}
\tau \rightarrow \infty
\begin{cases}
    \pi_A(i)=\pi_A(1)=\dfrac{(N-1)(\alpha-\beta)}{N-1} \rightarrow f_i=(\alpha-\beta)\\
    \pi_B(i)=\pi_B(1)=\dfrac{\alpha}{N-1} \rightarrow  g_i=\dfrac{\alpha}{N-1} \hspace{15pt}
\end{cases}
r=\dfrac{f_i}{g_i}=\dfrac{(\alpha-\beta)(N-1)}{\alpha}.
\end{split}
\label{eqcofi11}
\end{equation*}

Therefore:

\begin{equation}
    \Phi_{1,N} = \dfrac{1}{1 + \sum\limits_{k=1}^{N-1} \prod\limits_{j=1}^{k} \dfrac{1}{r}}=\dfrac{1-\dfrac{1}{r}}{1-{(\dfrac{1}{r})^N}}=\dfrac{1-\dfrac{\alpha}{(\alpha-\beta)(N-1)}}{1-{(\dfrac{\alpha}{(\alpha-\beta)(N-1)})}^N}.
    \label{eqcofi22}
\end{equation}

Based on this equation, when $\alpha$ and $\beta$ are held constant, for large enough $\tau$, $\Phi_{1,N}$ increases as $N$ increases, as shown in panel (a). In panel (b), we see that for N=4, the curves for ($\alpha=6$, $\beta=3$) and ($\alpha=10$, $\beta=5$), which both have $\frac{\alpha}{\beta}=2$, exhibit similar behaviour. This occurs because according to $U_2$ and Equations \eqref{1} and \eqref{efitnessfunction} when $N$ is held constant and $w=1$, the ratio $\frac{f_i}{g_i}$ remains the same for all values of $\alpha$ and $\beta$ as long as the ratio $\frac{\alpha}{\beta}$ is held constant. This leads to similar behaviour concerning variations in the time delay. Panel (c) illustrates how the population size impacts the fixation probability across three values of the time delay. Initially, increasing the population size reduces the probability of fixation regardless of the time delay. However, for sufficiently large populations, further increases in population size increase the probability of fixation. We observe differing patterns at small versus big population sizes. In small populations, longer time delays correspond to higher fixation probabilities for mutants. Conversely, in large populations, longer delays diminish the fixation probability.

In the snowdrift game, the impact of time delays on the fixation time is the reverse of what we saw in the Stag-Hunt game as shown in Fig. \ref{figsdti}. With a constant population size and payoff matrix values, increasing the time delay first lowers the fixation time, which then rises as the delay continues increasing. As before, for a large enough delay $\tau$, the fixation time converges to a constant value that is less than the fixation time when there is no delay.

\subsubsection{Prisoner's Dilemma game}
Another interesting case is the Prisoner's Dilemma. This game is characterized by the payoff relationship $c > a > d > b $ \cite{axelrod1981evolution,poundstone1993prisoner}. In such a setting the defect strategy D is a dominant strategy, so that for a rational player it is always optimal to play strategy D irrespective of the other player's choice. The standard Prisoner's Dilemma offers players the option of defecting or cooperating; for mutual cooperation, two interacting players are offered a reward, $a=R$, and for mutual defection, $d=P$. In this scenario, if one player cooperates while the other defects, then the cooperator would receive the sucker's payoff $b=S$, and the defector would receive the temptation-to-defect payoff $c=T$. Here we consider payoff matrices of the form:

\begin{center}
\begin{tabular}{ c c c c }
 &      & C & D \\ 
    & C & $\alpha$ & $\beta$ \\
$U_3$ = & & &     \\  
    & D & $\alpha+\beta$ & $1-\beta$    
\end{tabular}
\end{center}
where $\alpha>\beta$ and $0<\beta<0.5$.

\begin{figure}
{\includegraphics[scale=0.2]{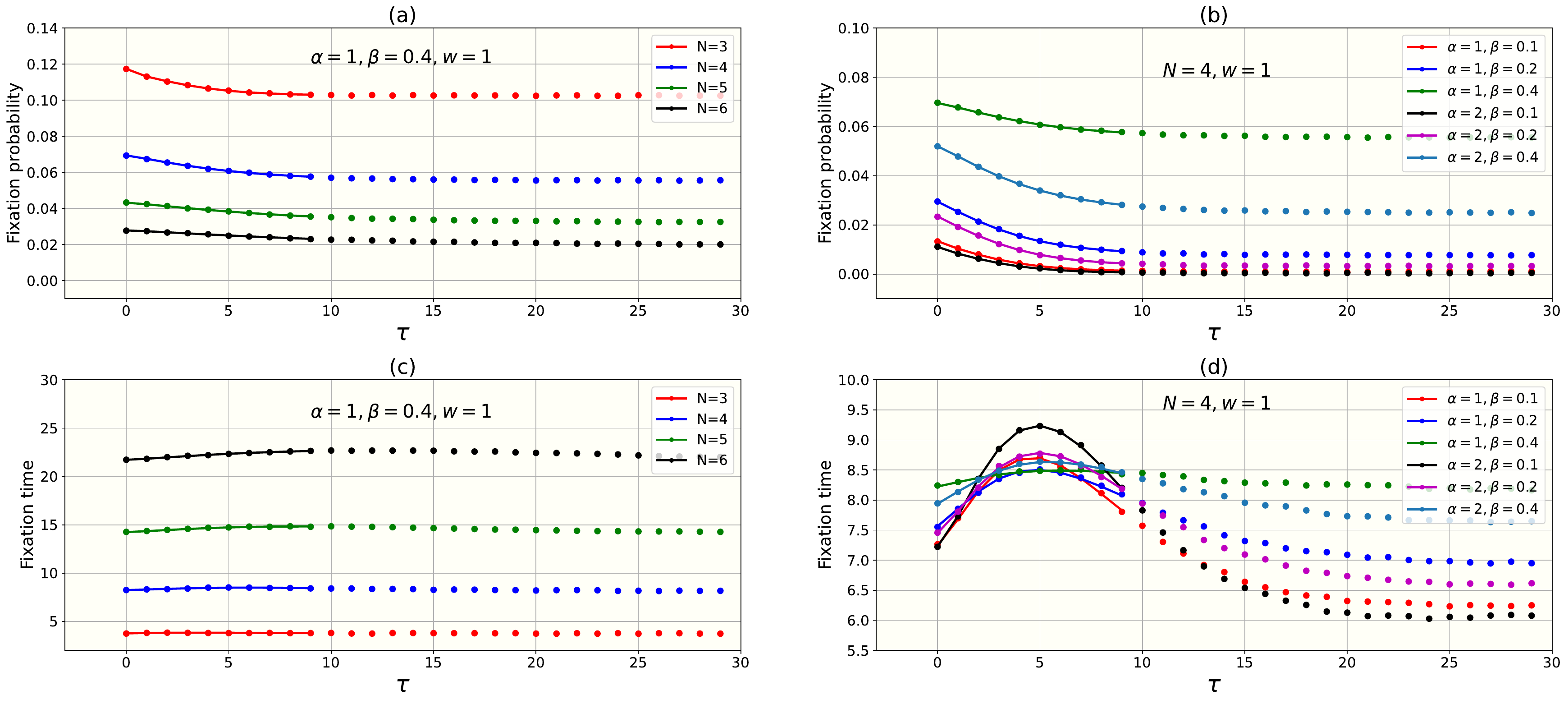}}
\centering
\caption{The effect of the time delay on the fixation probability and time in the Prisoner's Dilemma game ($U_3$) according to numerical solutions (lines) and computer simulations (dots) where the initial mutant is a cooperator. Panels (a) and (b) show the effect of time delays on the fixation probability and panels (c) and (d) show the effect of time delays on the fixation time for different network sizes and payoff values.} 
\label{figpdC}
\end{figure}

\begin{figure}
{\includegraphics[scale=0.2]{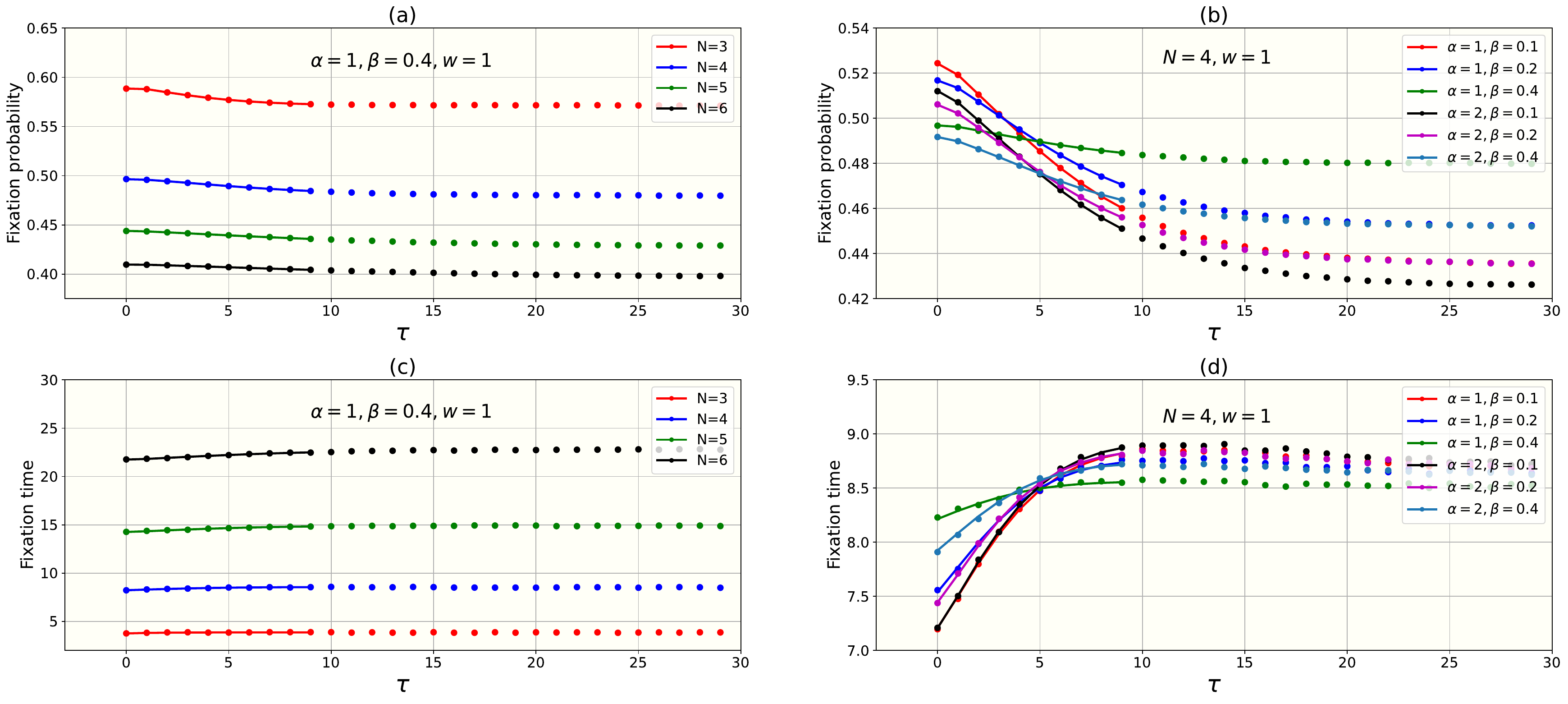}}
\centering
\caption{The effect of time delays on the fixation probability and time in the Prisoner's Dilemma game ($U_3$) according to numerical solutions (lines) and computer simulations (dots) where the initial mutant is a defector. Panels (a) and (b) show the effect of time delays on the fixation probability and panels (c) and (d) show the effect of time delay on fixation time for different network sizes and payoff values.} 
\label{figpdD}
\end{figure}

We examine two scenarios in this game. Fig.\ref{figpdC} and Fig.\ref{figpdD} depict how the fixation probability and time vary with changes in the time delay, for different population sizes and payoff matrices where the initial mutant is cooperator and defector respectively.  As expected, the fixation probability is much higher when the initial mutant is a defector rather than a cooperator under the same conditions since defectors, on average, obtain higher payoffs than cooperators when playing against both cooperators and defectors. In both scenarios, increasing the time delay leads to small changes in the fixation probability and time. However, larger time delays tend to decrease the fixation probability. We also observe a small increase followed by a small decrease in the fixation time as the time delay grows. This observation is clearer in the case where the initial mutant is a cooperator. For sufficiently large time delays, where the fitness depends only on the initial state, the constant fitness values ($r=\frac{f_1}{g_1}$) are $\frac{(N-1)\beta}{\alpha+N-2-(N-3)\beta}$ for an initial mutant with the cooperator strategy and $\frac{(n-1)(\alpha+\beta)}{\beta+(N-2)\alpha}$ for an initial mutant with the defector strategy (when $w=1$).  

\subsection{Conditional sojourn time}
The impact of time delays on the conditional sojourn time will be examined in this section.  Conditional sojourn time refers to the duration a process spends in a particular state before fixation. The conditional sojourn time $j$ starting from state $i$ refers to the number of times a process is in a transition state with $j$ mutants before fixation occurs\cite{grinstead1997introduction,hindersin2014counterintuitive}. Understanding conditional sojourn times is crucial for examining how traits or genotypes evolve within populations over time. When we introduce a time delay into models, we can analyze how the sojourn times in each state before fixation change. Specifically, the fixation time starts from $1$ mutant which is the sum of the sojourn times across all states before fixation occurs. Therefore, by studying how time delays affect sojourn times, we can better comprehend their impact on overall fixation times. 

For a process without time delays, when the process starts with $i$ mutants, the conditional sojourn time in the state with $j$ is $\dfrac{\Phi_{j,N}}{\Phi_{i,N}}F_{ij}$. With time delays, each state with $i_0=j$ mutants may have different historical components. To find the conditional sojourn time associated with a state having $j$ mutants, we sum over all $s_m$ where $i_0=j$. Using the fundamental matrix $F^\tau$, the conditional sojourn time for a state with $i_0=j$ mutants starting from state $s_m$ is defined as:

\begin{equation}
    \text{Conditional sojourn time (j)}=\sum_{s_l=\{i_\tau,...,i_0\}, i_0=j}^{S_T^\tau}\dfrac{\phi_{s_l,N}^\tau}{\phi_{s_m,N}^\tau} F_{s_m,s_l}^\tau .
\end{equation}

\begin{figure}
    \centering
    \includegraphics[scale=0.3]{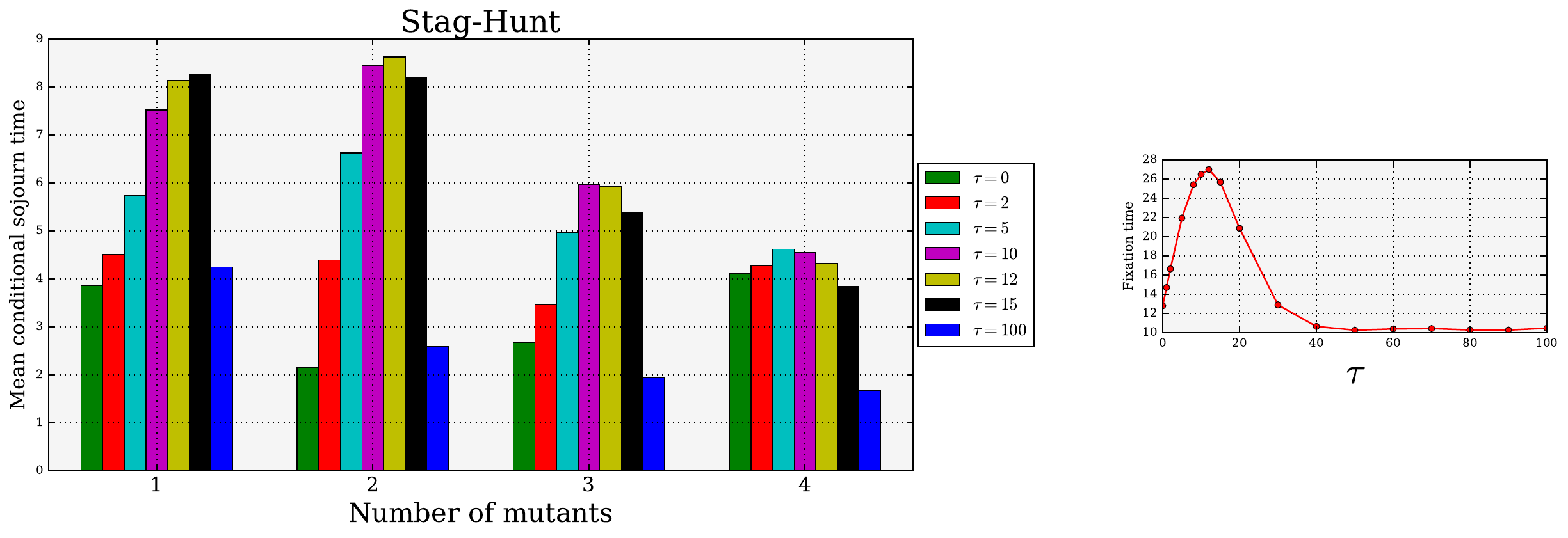}
    \caption{Conditional sojourn times for the Stag-Hunt game in a population of size 5 for different values of the time delay.}
    \label{figshsj}
\end{figure}

\begin{figure}
    \centering
    \includegraphics[scale=0.3]{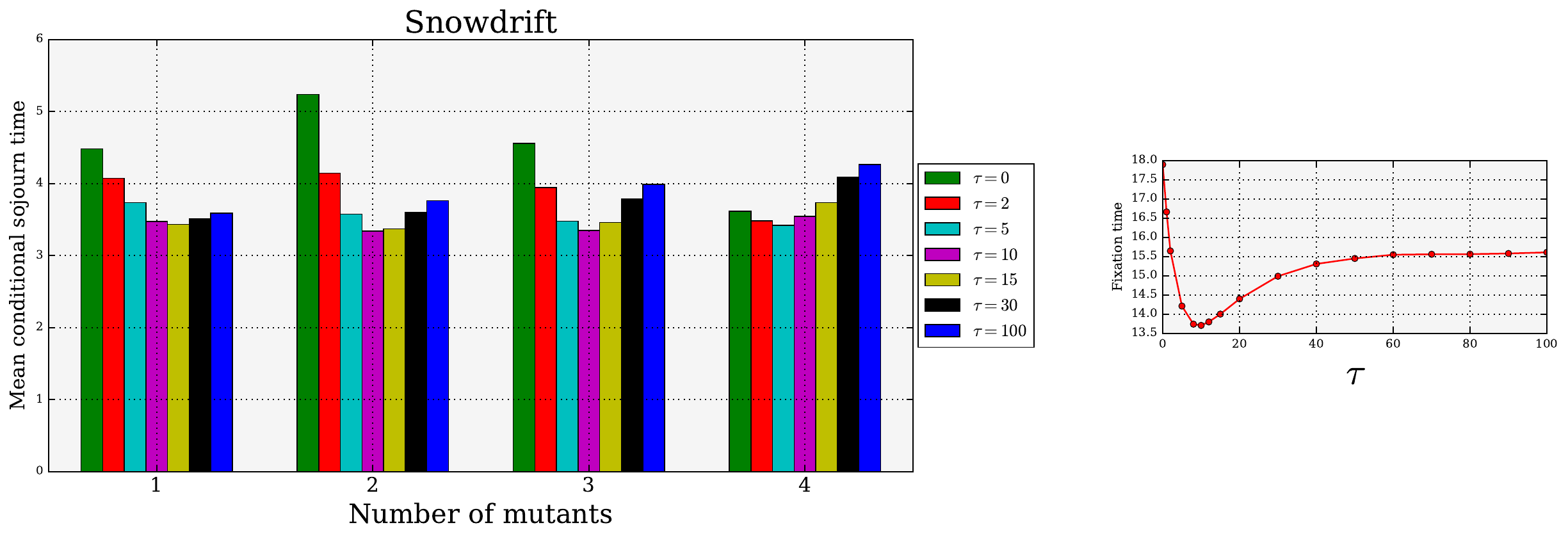}
    \caption{Conditional sojourn time for the Snowdrift game in a population of size of 5 for different values of the time delay.}
    \label{figsdsj}
\end{figure}

Fig \ref{figshsj} and Fig. \ref{figsdsj} show the mean conditional sojourn time for the Stag-Hunt and Snowdrift games, respectively, in a population of size 5. For the Stag-Hunt game, the payoff matrix $U_1$ is used with $\alpha=6$ and $\beta=0.2$. With no time delay, the smallest conditional sojourn time is when there are two mutants in the population. As the time delay increases to $12$, which corresponds to the largest fixation time as seen in the right panel, the conditional sojourn time for two mutants increases. On average, the conditional sojourn time at $\tau=12$ is higher than for other time delays, matching the highest fixation time. This suggests $\tau=12$ is a critical value where the process spends more time in transition states for $N=5$. For a sufficiently large delay (here $\tau=100$), where the fixation time converges to a constant, the conditional sojourn time decreases as the number of mutants increases, so that the process spends a small fraction of time in states with more mutants. Compared to the case without time delays, $\tau=100$ has a much smaller conditional sojourn time in states with four mutants. In other states, the conditional sojourn times are almost the same for $\tau=0$ and $\tau=100$. This leads to a lower fixation time for $\tau=100$ compared to $\tau=0$.

In the Snowdrift game, the results are the reverse of those for the Stag-Hunt. Using the $U_2$ payoff matrix with $\alpha=6$ and $\beta=4$, the largest conditional sojourn time on average occurs when there are two mutants. This time decreases as the time delay increases from $0$ until the critical delay of $10$ and then increases again. On average, $\tau=10$ has the smallest conditional sojourn time in all states, matching the lowest fixation time. In contrast to the Stag-Hunt game, for a sufficiently large delay, the conditional sojourn time increases as the number of mutants increases.

\section{Discussion}
Understanding fixation processes is vital for analyzing evolutionary games in finite populations. Two important metrics - fixation probability and time - have received significant research attention across various evolutionary game models\cite{de2019fixation,traulsen2006stochastic,traulsen2007pairwise,broom2010evolutionary,allen2021fixation,fudenberg2006evolutionary,antal2006fixation,sample2017limits,pires2022more,mobilia2010fixation}. Most previous models frequently assume that fitness instantaneously adjusts according to the current population state and payoffs\cite{de2019fixation,traulsen2006stochastic,traulsen2007pairwise,broom2010evolutionary,allen2021fixation,fudenberg2006evolutionary,antal2006fixation,sample2017limits,pires2022more,mobilia2010fixation,assaf2010large,czuppon2018fixation,whitlock2003fixation,ashcroft2014fixation,ohtsuki2006simple,broom2008analysis,allen2014games,lieberman2005evolutionary,shakarian2012review,hindersin2016exact,shakarian2013novel,hindersin2014counterintuitive,hajihashemi2019fixation,mohamadichamgavi2023effect}. This assumption of immediate fitness determination may not reflect reality and most of real phenomena exhibit temporal delays. Therefore we have  investigated the intricate influence of time delays on mutant fixation in evolutionary games.

Previously, researchers have studied the effect of time delays in deterministic replicator equation models\cite{yi1997effect,alboszta2004stability,mikekisz2021evolution,iijima2012delayed,ben2018discrete,wang2023evolutionary,miekisz2011stochasticity,wesson2016hopf,hu2019stability,wettergren2023replicator,bodnar2020three}. They have been incorporated in two different ways: first, by considering only the payoffs at time $t$ coming from time $t-\tau$ in the past\cite{yi1997effect,alboszta2004stability}; and second, by allowing individuals born in the past to replicate now based on past payoffs \cite{mikekisz2021evolution}. These studies have focused on how time delays impact the existence and stability of interior stationary states. Under the first approach, sufficiently long delays can generate limit cycle oscillations after a supercritical Hopf bifurcation\cite{yi1997effect,alboszta2004stability}. Under the second approach, strategy-dependent delays can shift the locations of stationary states to disfavour delayed strategies\cite{mikekisz2021evolution}. However, the impact of time delays on fixation dynamics, had remained an open question. 

Here we employed a well-mixed population model where individuals adhere to one of two strategies, with payoffs contingent upon their strategic choice and the prevalence of strategies\cite{taylor2004evolutionary}.  Initially, a solitary mutant individual with strategy A emerges amidst a population of residents employing strategy B. At each discrete time step, an individual reproduces in proportion to its payoff, randomly replacing another individual\cite{nowak2004emergence,taylor2004evolutionary}. In the absence of a time delay, reproductive fitness is determined solely by the number of mutants in the current state\cite{nowak2004emergence,ohtsuki2006simple,taylor2004evolutionary,pattni2017evolutionary,traulsen2006stochastic,hathcock2019fitness}. However, when a time delay $\tau$ is introduced, reproductive fitness becomes contingent upon the population state $\tau$ steps back in time.

Generally, both strategies' fitness depends on the frequencies of the strategies. When a mutant has relatively higher fitness in the population, it has a greater chance of becoming fixed. When we add a time delay to the model, it changes the relative fitness of mutants and residents because looking at past strategy frequencies can raise or lower fitness based on the payoff matrix. In the Stag-Hunt game, a mutant adopting strategy A experiences a greater reward when collaborating with fellow A mutants compared to interacting with B residents\cite{skyrms2004stag}. On the contrary, residents sticking to strategy B receive the same payoff regardless of whether they play with A or B strategies. As the A mutant population expands, they reap enhanced benefits by coordinating with each other, boosting their payoff (and fitness), consequently favoring mutant fixation. However, introducing a time delay disrupts this advantage by basing mutant reproduction on a past state where A mutants were less prevalent. This lagged payoff is lower than the real-time payoff, hindering A mutants' ability to reproduce and thrive. Notably, this time delay does not affect B residents' payoff. As a result, the effective payoff for A mutants diminishes, leading to a reduced fixation probability.

In contrast to the Stag-Hunt game, in the Snowdrift game, both A mutants and B residents attain higher payoffs from coordinating with A mutants\cite{broom2022game,doebeli2005models}. As the time delay increases, both strategies interact more with B residents, reducing their payoffs. However, A mutants maintain a superior effective payoff compared to B residents, enlarging the fitness ratio ($\frac{f_i}{g_i}$) beneficial to mutants. Thus, unlike the Stag-Hunt, lengthening the time delay in the Snowdrift game enhances A mutants' fixation prospects. In the Prisoner's Dilemma game\cite{axelrod1981evolution,poundstone1993prisoner}, the cooperator and defector strategies obtain higher payoffs when playing with a cooperator. When the initial mutant is a cooperator, increasing the time delay leads to a decrease in the effective payoff of both strategies for reproduction. However, the effective payoff for the cooperator from the past is smaller than for the defector, so the fixation probability decreases. When the initial mutant is a defector, although both strategies get more payoff when fitness comes from the past due to playing with more cooperators, the ratio of $\frac{f_i}{g_i}$ is smaller when there are more cooperators. So the time delay decreases the fixation probability regardless of the strategy of the initial mutant.  

In general, introducing a time delay alters the effective payoffs that both mutant and resident obtain for reproduction, impacting the fixation probability. A time delay tends to reduce the fixation probability if it reduces the relative payoff advantage of the mutant over the resident ($\frac{f_i}{g_i}$). It tends to increase the fixation probability if it enhances this comparative advantage. The key factor is the relative change in payoffs for mutant versus resident resulting from the delay. In summary, there are three scenarios where the time delay reduces the fixation probability:
\begin{itemize}
    \item If it decreases the mutant's payoff but increases the resident's payoff
    \item If it decreases both payoffs but the resident's proportionately less than the mutant's
    \item If it increases both payoffs but the resident's proportionately more than the mutant's
\end{itemize}

Conversely, there are scenarios where the time delay increases the fixation probability:
\begin{itemize}
    \item If it increases the mutant's payoff but decreases the resident's payoff
    \item If it increases both payoffs but the resident's proportionately less than the mutant's
    \item If it decreases both payoffs but the resident's proportionately more than the mutant's
\end{itemize}

We also have examined the joint effect of time delay and population size on fixation probability in stag-hunt and snowdrift games. Previous work showed population size can increase fixation probability in some games \cite{pires2022more}. Here, in stag-hunt games, increased population size decreases fixation probability regardless of time delay, although time delay affects smaller populations more. In contrast, in snowdrift games, increased population size increases fixation probability. However, time delay affects small and large populations differently: bigger delays help mutant fixation in small populations, while smaller delays help in large populations.

Fixation time has been less studied in past literature. This measurement can vary substantially depending on the model used and the structure of the population\cite{antal2006fixation,traulsen2009stochastic,hindersin2014counterintuitive}. In some cases, fascinating behavioral patterns may emerge related to fixation time. The effect of time delays on fixation times varies across the three studied games. In the Stag-Hunt game, as time delays initially increase, fixation times grow longer, peaking and then decreasing after a critical time delay point. Analyzing the conditional sojourn time (the duration the process stays in each state before fixation), we see the process lingers mostly in states with few or many mutants in this game. However, the process spends more time in intermediate states as delays rise until reaching a peak, then declining. In contrast, the Snowdrift game demonstrates the opposite pattern – fixation times decrease and then increase again after a critical delay point. Here, conditional sojourn times in intermediate states dramatically fall with longer delays until the minimum fixation time is reached. Meanwhile, time delays have little influence on fixation times in the Prisoner’s Dilemma game, though there is a slight increase then decrease similar to the Stag-Hunt game.

Our findings suggest that incorporating time delays into evolutionary game dynamics introduces novel and potentially significant effects. While our study focused on scenarios with uniform time delays for both mutant and resident strategies, further investigations are needed to explore cases where the time delay is strategy-dependent. Additionally, our exploration of the Bd update rule, where birth is influenced by fitness and time delay, highlights the potential of alternative update rules, such as birth-Death, death-Birth, Death-birth, and imitation, to unveil distinct time delay effects. Moreover, our analysis of well-mixed populations without structure provides a foundation for future studies in structured populations, including line graphs, star graphs, and more complex graphs such as scale-free and random graphs. Examining strategy-dependent time delays across structured populations represents a promising direction for future research on time delays in evolutionary games.

\section*{Acknowledgments}
This project has received funding from the European Union’s Horizon 2020 research and innovation program 
under the Marie Sk\l odowska-Curie grant agreement No 955708.



%
%
\bibliographystyle{RS}
\bibliography{main}


\end{document}